\documentclass[preprint,showpacs,12pt,nofootinbib]{revtex4}

\usepackage{amsmath,amssymb,amsfonts}
\usepackage{graphicx}
\usepackage{hhline}
\usepackage{bm}
\usepackage{amsmath}
\usepackage{epsfig}
\usepackage{graphicx}

\usepackage{tabularx}
\usepackage{lscape}
\usepackage{rotating}
\usepackage{pstricks}

\usepackage{footnote}
\usepackage{hyperref}

\usepackage{pdfpages}
\usepackage{color} 

\usepackage{enumitem}
\begin{document}

\title{Reflections on Chiral Symmetry within QCD}

\author{Anthony W Thomas}

\affiliation{CSSM and ARC Centre of Excellence for Dark Matter Particle Physics, \\
Department of Physics,\\ 
The University of Adelaide SA 5005 Australia} 

\begin{abstract}
That chiral symmetry is a crucial feature of the strong force was realized before the discovery of Quantum Chromodynamics. However, the full power it exerts on the structure of the nucleon became apparent only afterwards. We present a high-level and somewhat personal overview of its role in almost every aspect of proton structure, from its mass and spin to the asymmetry of its antimatter content and its strange quark content. The lessons learned from studying the proton are also vital with respect to the modern challenge of the nature of baryon excited states.

\end{abstract}

\maketitle

\section{Introduction}
With the passing of 50 years since the creation of Quantum Chromodynamics, QCD, with its myriad of successes, much has been written~\cite{Gross:2022hyw}. From the initial ideas of quarks~\cite{Gell-Mann:1962yej} and aces~\cite{Zweig:1964ruk} to the emergence of a local gauge theory of quarks and gluons at the hands of Fritzsch, Gell-Mann and Leutwyler~\cite{Fritzsch:1973pi}, it took a decade. I first heard of it at the International Conference on High Energy and Nuclear Physics (later called PANIC) in Los Alamos in 1974 when Gell-Mann gave an inspired after dinner talk about it. It was evident from his presentation that he considered it worthy of another Nobel Prize but that was awarded to Gross, Wilczek and Politzer~\cite{Politzer:1973fx,Gross:1973id} for demonstrating that QCD incorporated asymptotic freedom. 

Of course, the J/$\Psi$ was discovered in November 1974~\cite{E598:1974sol,SLAC-SP-017:1974ind} but I recall that in 1975 at CERN this particle was still an enigma. Influenced by the mystery of the $J/\Psi$, the Theory Division notice board in late 1975 had an announcement of "A New Resonance Discovered" in the Division. This was actually the dramatic rise in computer time consumed by my Faddeev calculations~\cite{Rinat:1976qg}. It took what might now seem an inordinate amount of time for the $J/\Psi$ to be identified as the 
$c-\bar{c}$ bound state we know and love, especially given the inspired work of Glashow, Illiopoulos and Maiani~\cite{Glashow:1970gm}, but this marked the point where quarks were accepted as real. For QCD itself, it was the discovery of three-jet events in $e^+-e^-$ annihilation at DESY in 1978-79~\cite{TASSO:1980lqw} that convinced the community that we had the correct gauge theory for the strong interaction.

At MIT, theorists had already constructed a  relativistic theory of free quarks confined in a volume called a bag, with a different vacuum structure inside and out~\cite{Chodos:1974je,Chodos:1974pn}. However, it was soon realized that a fundamental symmetry of QCD, namely chiral symmetry~\cite{Pagels:1974se}, was violated by the MIT bag. In particular, as confinement is modeled by an infinite mass term at the bag boundary, which is a Lorentz scalar, when a quark is reflected at the surface its helicity is flipped. This contradicts the chiral symmetry of QCD for massless quarks.

The resolution of this problem had its origin in the discovery of the partially conserved axial current, 
PCAC~\cite{Gell-Mann:1960mvl,Glashow:1967rx,Gell-Mann:1964hhf}, which was so widely used in the 1960's to describe pion-nucleon interactions~\cite{Gell-Mann:1960mvl,Adler:1964um,Weinberg:1966kf}. By coupling pions to the quarks at the surface of the bag, realizing chiral symmetry in the Goldstone mode, one could restore the symmetry. Early work by Chodos and Thorn, using this idea, was a semi-classical extension of the linear sigma model~\cite{Chodos:1975ix}.

The development of this idea into a quantitatively successful chiral model of hadron structure, the cloudy bag model (CBM), followed after discussions with Gerry Brown about his ''little bag''~\cite{Brown:1979ij}, as well as a somewhat scary flight from Houston to Denver en route to Seattle and Vancouver. As the plane from Houston circled for an hour over Denver, because of bad thunderstorms that rocked the aircraft, passengers were allowed to walk about (unimaginable now!) and Gerry Miller and I discussed the apparent contradiction in Brown's ideas. He wanted the pion to squeeze the bag to a tiny size in order to fit his ideas about the nucleon-nucleon force. But we realized that if this were so, the Chew-Low mechanism~\cite{Chew:1955zz} would generate a second $\Delta$ resonance, in addition to that given by the quark model.

The solution, involving a collaboration with Gerald Miller and Serge Th\'eberge,  was the first serious treatment of meson-baryon interactions in the context of chiral symmetry and the quark model~\cite{Miller:1979kg}. From this~\cite{Thomas:1982kv}, modern studies of the spectrum of baryon resonances in lattice QCD have evolved, but more on that later. In the CBM the high momentum cut-off, or form factor, at the $N \rightarrow \pi N$ and $N \rightarrow \pi \Delta$ vertices have their origin in the finite size of the source. Because both the $N$ and $\Delta$ are ground states, with their valence quarks in the lowest state, they have essentially the same size (bag radius). As a consequence, these form factors were equal, up to the respective coupling constants, which were also predicted. This led unambiguously to the conclusion that the $\Delta$ resonance was predominantly a three-quark state, with a small fraction of the width arising through the Chew-Low mechanism~\cite{Theberge:1980ye}. It also required that the size of the nucleon bag be not too small, with 0.8 to 1.0 fm preferred. With such a radius one can prove rigorously that the number of pions in the cloud around the bag is quite small~\cite{Dodd:1981ve}.

\begin{figure}[!ht]
\begin{center}
\includegraphics[width=0.8\columnwidth]{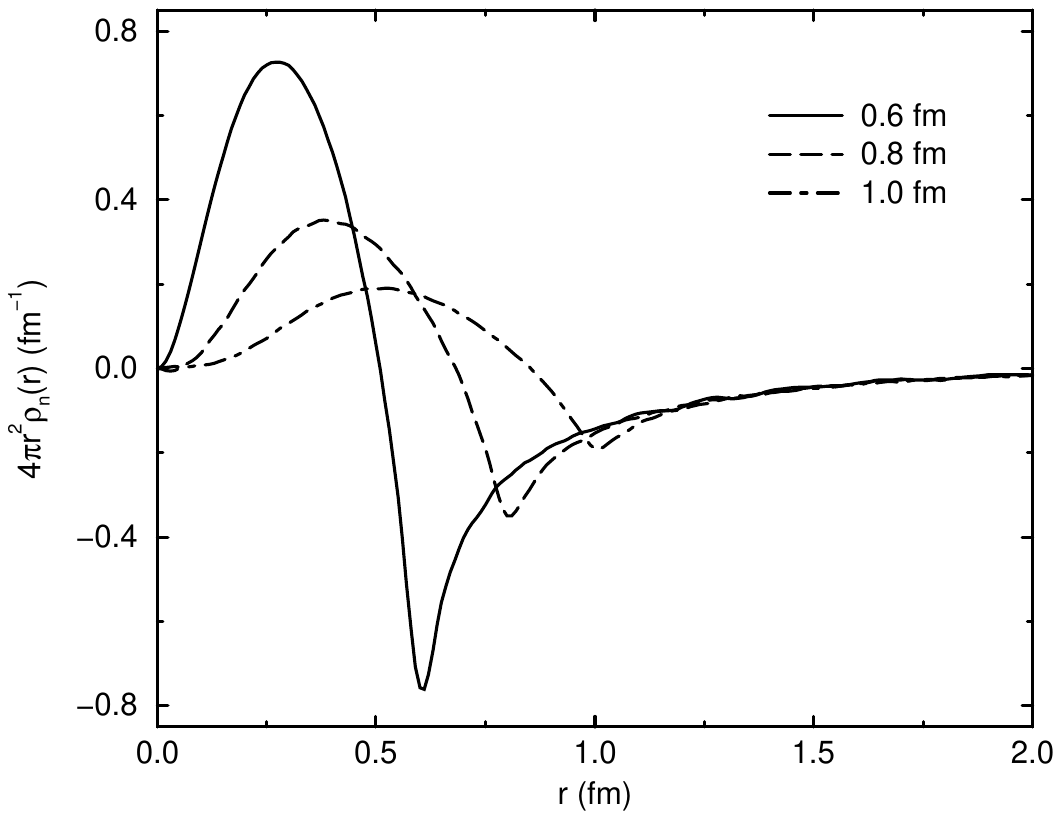}
\vspace*{-0.2cm}
\caption{Neutron charge density calculated in the simplest version of the CBM (with a sharp surface at the bag radius) for several choices of bag radius, R.
Note that the peak in the negative charge density always occurs very near the chosen value of R.  }
\label{fig:neutron}
\end{center}
\end{figure}

Another highlight of this approach, which was evident immediately, is that it naturally explained the charge distribution of the neutron. In the bag model the three-quark neutron has no charge distribution. However, the leading chiral component involves a negative pion cloud peaked at the bag surface and extending outside, while the core is a proton bag. This naturally implies a change in sign of the charge distribution in the region of the bag surface~\cite{Thomas:1981vc}. For the simplest version of the CBM, the neutron charge distribution is illustrated in Fig.~\ref{fig:neutron} for several bag radii. The study of the $\Delta$ resonance suggested that the radius should be between 0.8 and 1.0 fm, which is indeed where the charge distribution of the neutron changes sign~\cite{Hofstadter:1961zz,Galster:1971kv,Bartel:1973rf,Platchkov:1989ch}.~\footnote{There has been considerable discussion recently about the interpretation of the charge distribution as the Fourier transform of the electric form factor~\cite{Miller:2007uy}, suggesting that the infinite momentum frame should be preferred. However, the long range behavior involves low momentum, where relativistic corrections are not expected to be significant. We also note that Lorc\'{e} has shown that the Fourier transform in the Breit frame yields internal charge quasidensities in the rest frame of a localized target, without any relativistic correction~\cite{Lorce:2020onh}. }

With the pion field quantized, one could calculate the electroweak form factors of the entire baryon octet. While a bag of radius 1.0 fm yields an rms charge radius smaller than the experimental value, the pion cloud which extends well beyond the bag radius ensures agreement with experimental data~\cite{Thomas:1981vc,Theberge:1981pu,Theberge:1982xs,Zenczykowski:1982rt,Guichon:1982zk}. 

In the following we review more of the remarkable physics which can be understood in terms of the chiral structure of QCD.

\section{Anti-matter in the proton}
In late 1982, not long after I had arrived at CERN as a staff member in the Theory Division, on leave from TRIUMF, Erwin Gabathuler wandered into my office seeking insight into the remarkable data that the European Muon Collaboration (EMC) had collected. That data demonstrated a remarkable change in the structure function of a nucleus compared with a free nucleon, with the valence quarks losing momentum~\cite{EuropeanMuon:1983wih,EuropeanMuon:1992pyr}. This famous result is known as the EMC effect, which has been the subject of extensive experimental investigation~\cite{BCDMS:1985dor,Arrington:2021vuu}, 
is still a source of controversy~\cite{Geesaman:1995yd,Hen:2016kwk,Thomas:2018kcx}.

Magda Ericson, Chris Llewellyn Smith and I began work on the possibility that an enhancement of the pion field per nucleon in a nucleus might explain the effect. Magda and I published our result~\cite{Ericson:1983um} back-to-back with Chris~\cite{LlewellynSmith:1983vzz}, even though the work had been carried out together. 

The reason for this is fascinating. At that time, Chris, like the entire high-energy community, refused to take seriously the idea of deep-inelastic scattering (DIS) from the virtual pion.  For him the pion cloud merely served as a device to steal momentum from the nucleons in the nucleus. This may seem hard to believe now, with the ''Sullivan process''~\cite{Sullivan:1971kd} widely cited, but in the early 1980s almost no-one took it seriously.

On the other hand, Magda and I did take the concept of DIS from the pion cloud seriously. As a further consequence, within the CBM I showed that the pion cloud of the nucleon is necessarily larger than the kaon cloud and as a consequence the strange sea is suppressed in comparison with the non-strange sea~\cite{Thomas:1983fh}. However, the most remarkable result in that work was the {\em prediction} that the anti-down sea of the proton should be considerably larger than the anti-up sea. I was extremely nervous to publish this result, given the lack of tolerance in the high energy community, but the consequence of chiral symmetry was unavoidable. The degree that this was outside the mainstream at the time is that this paper was hardly cited in the first few years after its publication. Indeed, it was a decade before a measurement of the Gottfried sum rule established that this predicted excess of anti-down quarks was indeed correct~\cite{NewMuon:1991hlj}.

This asymmetry in the anti-matter content of the nucleon sea has been the object of intense theoretical and experimental study for over 30 years. On the experimental side, the Drell-Yan measurements at Fermilab have slowly revealed the shape of the $\bar{d}$ excess over $\bar{u}$~\cite{NuSea:2001idv,FNALE906:2022xdu}, which are consistent with early predictions of the CBM approach~\cite{Melnitchouk:1991ui,Melnitchouk:1998rv}.

\subsection{Asymmetry in the strange sea}
With the CBM extended to chiral SU(3), the experience with the $\bar{d} - \bar{u}$ asymmetry suggested that the emission of a kaon in the process $N\rightarrow K Y$, with $Y$ a hyperon, would also lead to an asymmetry in the strange sea~\cite{Signal:1987gz}. This is because the strange quark would reside in a hyperon, while the anti-strange quark resides in the kaon. A second consequence is that the larger mass of the kaon means that the strange sea will be smaller than the light quark sea. For recent calculations of these distributions we refer to Refs.~\cite{Wang:2016ndh,Wang:2016eoq}.

It is important to realize that, while the insights provided by models such as the CBM are important,  chiral symmetry also makes {\em model independent} predictions. In particular, it is a model independent consequence of chiral symmetry that the shape of $s(x)$ and $\bar{s}(x)$ {\em must} be different~\cite{Thomas:2000ny}. The proof relies on the fact that only Goldstone boson loops give rise to non-analytic behavior as a function of quark mass. As shown in Ref.~\cite{Thomas:2000ny}, the leading non-analytic (LNA) terms (i.e., those involving the lowest power of quark mass -- strange quark mass in this case) are different for the moments of $s(x)$ and $\bar{s}(x)$. Indeed, it is a rigorous result of chiral symmetry that whereas the LNA behavior of the n'th moment of $\bar{s}$ is $m_K^{2n+2} \ln m_K^2$,  for the n'th moment of $s$ the behavior is $m_K^2 \ln m_K^2$, and hence the distributions {\em cannot as a matter of principle} be the same.

\begin{figure}[!h]
\begin{center}
\includegraphics[width=1.0\columnwidth]{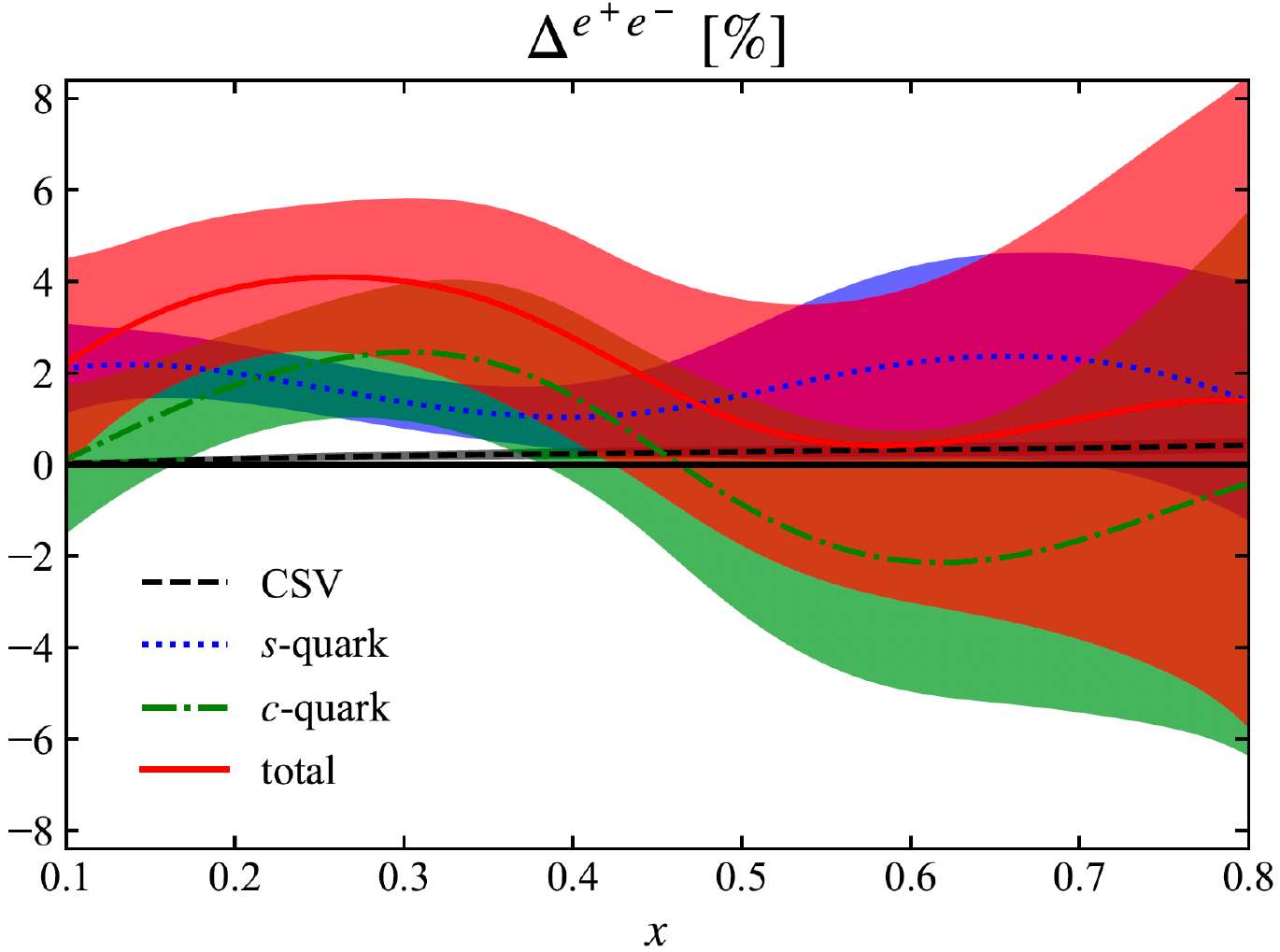}
\vspace*{-0.2cm}
\caption{Corrections to the positron-electron asymmetry in DIS from the deuteron at $Q^2 = 10\ {\rm GeV}^2$. The uncertainties arising from the lack of knowledge of the C-odd strange and charm PDFs seriously reduce the limits one can set on the scale of new BSM physics. (From Ref.~\cite{Wang:2024mll}.) }
\label{fig:Delta_C3q_10GeV2_nnpdf}
\end{center}
\end{figure}

Sadly, after almost 50 years of studying parton distribution functions (PDFs) our knowledge of the strange and anti-strange PDFs is still poor~\cite{NNPDF:2021njg,Ablat:2024muy,Melnitchouk:2021djs}. This is especially significant because an accurate knowledge of them is crucial to precision tests of physics beyond the Standard Model using parity violating electron scattering~\cite{Wang:2024mll}. As an example, in Fig.~\ref{fig:Delta_C3q_10GeV2_nnpdf} we show the correction to the predicted asymmetry between $e^+$ and $e^-$ DIS from the deuteron arising from charge symmetry violation~\cite{Rodionov:1994cg,Sather:1991je}, and the C-odd strange and charm PDFs. This particular observable is sensitive to the combination of axial couplings of the leptons and quarks~\cite{Zheng:2021hcf}.

 These uncertainties in the phenomenological strange quark PDFs extracted from DIS data also mean that, for the moment, we cannot test the predictions for $s(x) \, - \,\bar{s}(x)$ based upon chiral symmetry.

\section{QCD predictions as a function of quark mass}
In the early days of QCD it was crucial to test that it was indeed the correct theory of the strong interaction. The first challenge was to use the remarkable development of lattice QCD~\cite{Wilson:1974sk} to calculate the mass of the nucleon. In that regard, perhaps I may be forgiven an anecdote from my time at CERN as a research fellow in 1976. Sasha Migdal had been allowed out of Russia to come to CERN for a year and shared an office with me. I recall that he was extremely excited to have acquired a Hewlett-Packard programmable calculator, with the famous reverse-Polish notation. He was excited because he believed at the time that this would give him the computing power, after his return to Moscow, to calculate the mass of the nucleon in lattice QCD.  

Although realistic lattice QCD calculations required more than a handheld calculator, modern calculations, corrected for finite volume and finite lattice spacing, do reproduce the mass of the baryon octet very accurately~\cite{BMW:2008jgk,BMW:2013mpk}. Along the way, numerical issues meant that early calculations were performed at larger quark masses than Nature provides. In order to compare with experimental data it was necessary to extrapolate from the large mass region to the physical mass. 

Chiral perturbation theory~\cite{Gasser:1983yg,Leutwyler:1993iq,Meissner:1993ah,Ecker:1994gg,Pich:1995bw} was the natural choice, but was challenged by the fact that its radius of convergence is typically a pion mass around 300 MeV (or a quark mass as low as 10-20 MeV). Every time a higher power was added to the perturbative expansion, the behavior of different nucleon observables at large pion mass, or equivalently large quark mass (as $m_\pi^2 \propto m_q$), diverged in a different way~\cite{Young:2002ib}. However, an examination of all nucleon properties calculated in lattice QCD revealed that they all behave smoothly at large pion masses. Indeed, they look very much like one would expect in a constituent quark model.

\begin{figure}[t]
	\centering
	\includegraphics[width=\columnwidth]{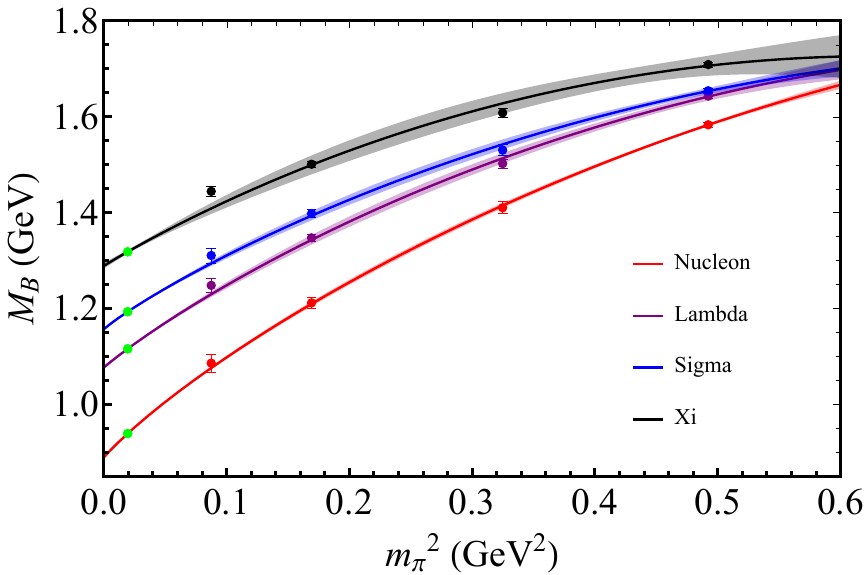}
	\caption{Fits to the PACS-CS octet baryon mass results in the covariant formalism at $\Lambda_{(\text{Cov})}=1.0$ GeV, with the inclusion of the physical points (green dots). The data points are finite-volume corrected. The extrapolation bands and errors on individual data points are purely statistical -- from Ref.~\cite{Owa:2023tbk}.}
\label{fig:04}
\end{figure}

The key to understanding this was already evident in the CBM. Because the source of the pion field has a finite size, characterized by the bag radius in that model, there are form factors at the pion-baryon vertices that suppress high-momentum pions. As a consequence, they also suppress pion loops at high pion mass~\cite{Leinweber:1999ig,Donoghue:1998bs}. The extrapolation method which exploited this feature of QCD was called finite range regularization (FRR). Figure~\ref{fig:04} illustrates a recent chiral fit, from Ref.~\cite{Owa:2023tbk}, to the data for the masses of the octet baryons from the PACS-CS Collaboration~~\cite{PACSCSdata}.

Fits such as those shown in Fig.~\ref{fig:04} can be used to extract the baryon sigma commutators, which are a direct measure of the contribution of the quark mass term in the QCD Lagrangian to the baryon masses~\cite{Alarcon:2021dlz}. The contribution of the $u$ and $d$ quarks to the nucleon mass, known as $\sigma_{\pi N}$, is particularly interesting as phenomenological analysis yields values of order 60 MeV~\cite{Hoferichter:2015dsa,Hoferichter:2023ptl,RuizdeElvira:2017stg,Yang:2015uis}, with error estimates ranging from 2 to 9 MeV, whereas the latest FLAG review of lattice results quotes a value around 42 MeV~\cite{FlavourLatticeAveragingGroupFLAG:2024oxs,Bali:2016lvx}. The recent chiral analysis of Owa {\em et al.}~\cite{Owa:2023tbk}, using the lattice results of the CLS Collaboration~\cite{Ottnad:2022axz}, gave a value around 52 MeV, with an error of order 4 MeV, which is not incompatible with the phenomenological extractions.

Another important result from the study of the nucleon mass within chiral SU(3) was the ability to extract the strange sigma commutator, $\sigma_s \, = \, m_s \, \langle N| \,  \bar{s} s \, |N \rangle$. This is especially important in the analysis of direct searches for dark matter in the context of supersymmetry. The neutralino-nucleon cross section is quite sensitive to $\sigma_s$ and it turned out that using FRR to analyze lattice data as a function of the strange quark mass, $m_s$, gave a value roughly an order of magnitude smaller than had earlier been believed~\cite{Ellis:2008hf,Giedt:2009mr}; i.e., between 20 and 60 MeV~\cite{Young:2009zb,Shanahan:2012wh}, rather than 300 MeV~\cite{Nelson:1987dg}.

Although the extrapolation of lattice results from a large pion mass to the physical value, in order to compare with experiment, was important, apart from the study of the $\sigma$ commutator, it is of less interest now that calculations at the physical mass are possible. Far more important now is the realization~\cite{Thomas:2002sj} that lattice QCD still provides the behavior of QCD even if the pion mass is not physical. This offers remarkable insights into hadron structure, especially when it comes to understanding the baryon spectrum. 

\begin{figure}[t]
\includegraphics[width=0.9\columnwidth,angle=0]{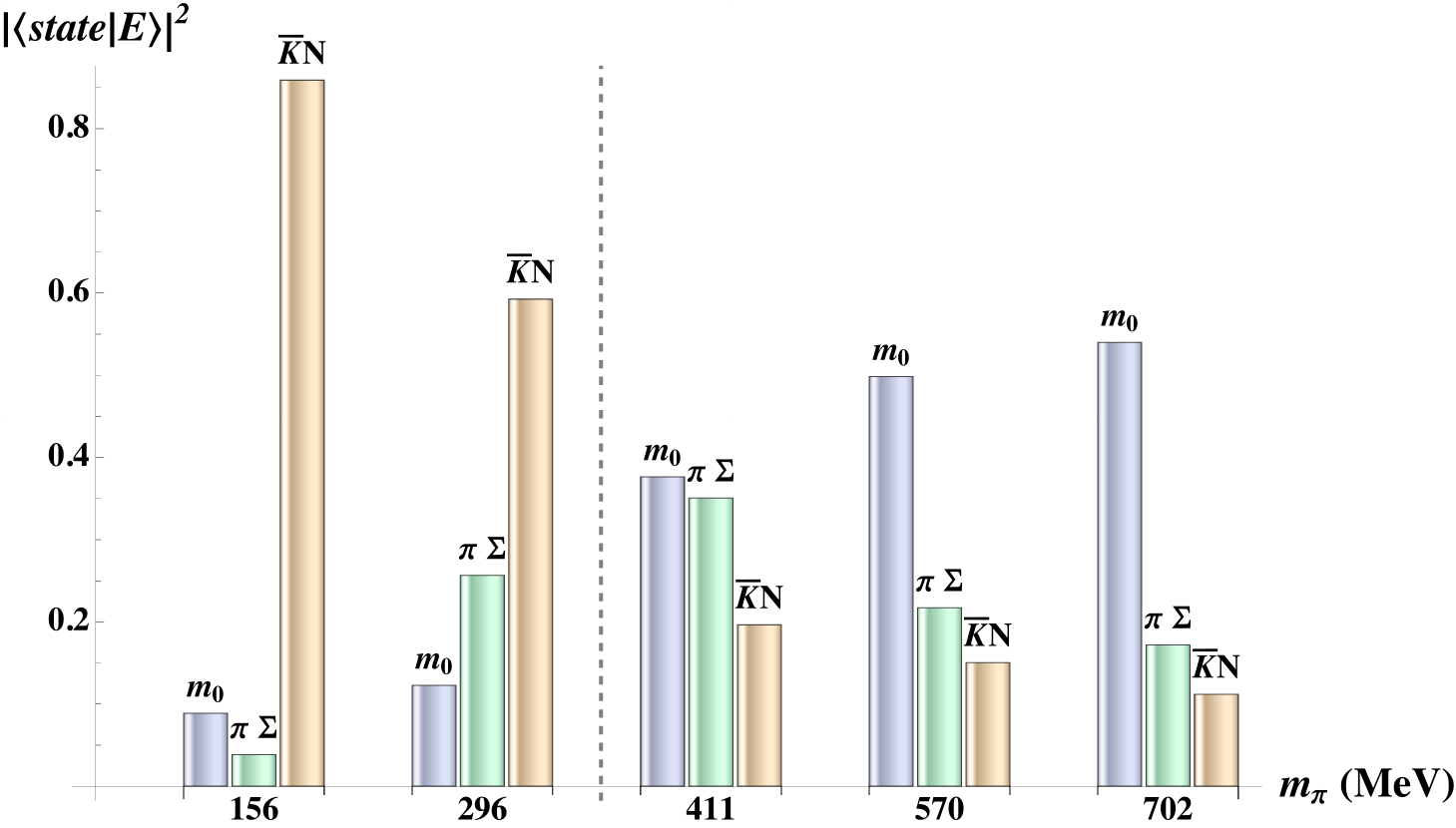}
\vspace*{-8pt}
\caption{The overlap of the basis state, $| {\it state} \rangle$, with
  the energy eigenstate $| E \rangle$ for the $\Lambda(1405)$,
  illustrating the composition of the $\Lambda(1405)$ as a function of
  pion mass.  Basis states include the single particle 
  state, denoted by $m_0,$ and the two-particle states
  $\pi\Sigma$ and $\overline{K}N$.  A sum over all two-particle
  momentum states is done in reporting the probability for the
  two-particle channels.  Pion masses are indicated on the $x$ axis
  with the vertical dashed line separating the first state for the
  heaviest three masses from the second state for the lightest two
  masses. (Reproduced with permission from Physical Review Letters, Ref.~\cite{Hall:2014uca}, published by the American Physical Society, 2015.).
\vspace*{-10pt}
\label{fig:Hist}}
\end{figure}

\begin{figure}[t]
\includegraphics[width=0.98\columnwidth]{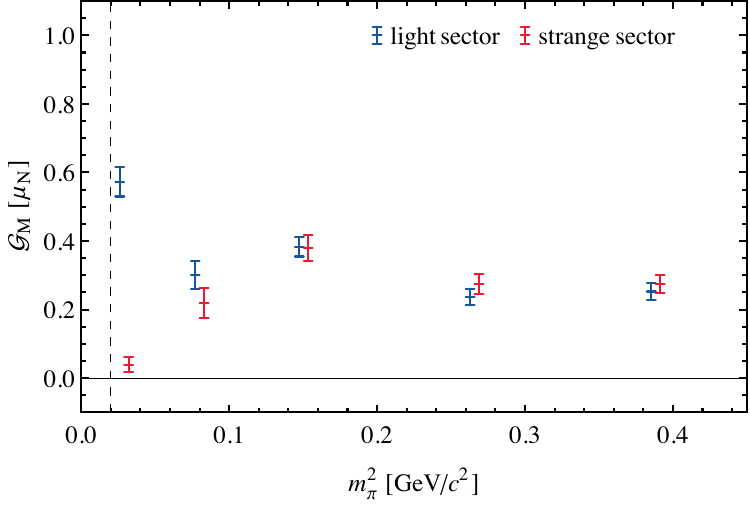}
\vspace*{-8pt}
\caption{The light ($u$ or $d$) and strange ($s$) quark contributions
  to the magnetic form factor of the $\Lambda(1405)$ at $Q^2 \simeq
  0.16$ GeV${}^2/c^2$ are presented as a function of the light $u$ and
  $d$ quark masses, indicated by the squared pion mass, $m_\pi^2$.
  Sector contributions are for single quarks of unit charge.  The
  vertical dashed line indicates the physical pion mass. (Reproduced with permission from Physical Review D, Ref.~\cite{Hall:2016kou}, published by the American Physical Society, 2017.)
\vspace*{-10pt}
\label{fig:GM}}
\end{figure}

Perhaps the most outstanding example of this is the $\Lambda(1405)$, which was originally suggested by Dalitz and Tuan~\cite{Dalitz:1960du} to be a $\bar{K} N$ bound state; a suggestion supported by the first chiral calculation after the discovery of QCD~\cite{Veit:1984jr}, using the volume coupled version of the CBM~\cite{Thomas:1981ps} which incorporates the Weinberg-Tomozawa relation~\cite{Weinberg:1966kf,Tomozawa:1980rc}. Using Hamiltonian effective field theory~\cite{Hall:2013qba}, one can explicitly see the evolution in character of this famous resonance as it changes from a three-quark state to a $\Bar{K} N$ bound state as the pion mass varies from 600 MeV to the physical value~\cite{Hall:2014uca,Hall:2016kou} -- see Fig.~\ref{fig:Hist}. As further evidence for this interpretation, we see in Fig.~\ref{fig:GM} that as we approach the chiral SU(2) limit the contribution of strange quarks to the magnetic moment of the $\Lambda(1405)$ vanishes. This is exactly what one expects if the strange quark is bound in a spin zero meson (the $\bar{K}$) with angular momentum zero.

\subsection{Strangeness content of the nucleon}
One of the fundamental tests of QED was the Lamb shift. For QCD the analogue is the strange quark contribution to the magnetic moment of the proton, because this originates entirely through virtual sea-quark loops. It is therefore vital to check that QCD does indeed predict the observed contributions to the proton electromagnetic properties from strange quarks.

In the early years of this century, this was also a hot topic because a popular resolution of the so-called ''spin crisis''~\cite{Veneziano:1989ei,Leader:1988vd}, corresponding to a significant violation of the Ellis-Jaffe sum rule~\cite{Ellis:1973kp}, discovered at CERN~\cite{EuropeanMuon:1987isl},  involved a significant enhancement of the strange sea. In addition, the feasibility of precise measurements of the parity violating asymmetry~\cite{Erler:2013xha} in polarized electron scattering at Jefferson Lab and Mainz offered an experimental method to test this proposal~\cite{Kaplan:1988ku}. 

In the period when experiments to measure the strange electric and magnetic form factors were underway, lattice QCD was unable to directly determine them. Furthermore, the calculations of individual up and down contributions were limited to relatively large quark masses in quenched QCD. In spite of these difficulties, using the insights into the chiral behavior of nucleon properties described earlier and handled quantitatively using FRR, it proved possible to unquench the lattice results~\cite{Leinweber:2002qb} and extrapolate them to the physical light quark masses to extract the strange magnetic moment and strange electric radius of the proton. The results obtained, namely $G_M^s = -0.046 \pm 0.019 \mu_N \,$~\cite{Leinweber:2004tc} and $G_E^s(0.1 {\rm GeV}^2) \, = \, +0.001 \pm 0.004 \pm 0.004 \, $~\cite{Leinweber:2006ug}, were in excellent agreement with the experimental results when they were published~\cite{HAPPEX:2006oqy,G0:2009wvv,HAPPEX:2011xlw} a few years later, thus confirming that QCD indeed satisfies the ''Lamb shift test''. Of course, recently lattice studies have been possible at physical quark masses~\cite{Alexandrou:2019olr} and these are in excellent agreement with the 2005-6 calculations using FRR, with a level of precision roughly a factor of two better.

\section{The ''Spin Crisis''}
In the late 1980s, experiments at CERN and elsewhere~\cite{EuropeanMuon:1987isl,Hughes:1988vh} established that the fraction of spin carried by the quarks in the proton was of the order one third of its total spin. Many explanations have been offered, which are now known to be incorrect, including the possibility of a large polarized strange quark sea (discussed in the previous section) and a large amount of spin carried by gluons~\cite{Jaffe:1987sx,Carlitz:1988ab,Bodwin:1989nz,Efremov:1989sn,Shore:1990zu,Bass:1991yx}.

A very natural explanation arises within the framework of chiral symmetry. This is the same physics which describes the neutron charge radius and the anti-matter asymmetry in the proton. In particular, when a proton emits a pion there is a high probability that its spin flips and the proton spin in this Fock state is carried as orbital angular momentum of the pion~\cite{Schreiber:1988uw}. Because the pion has spin zero there is no contribution to the fraction of quark spin from the pion cloud.

There is no doubt that a major piece of the explanation of the so-called spin crisis is this transfer of quark spin to pion orbital angular momentum. Certainly, gluon spin~\cite{Bazilevsky:2016itl,deFlorian:2014yva,STAR:2021mqa,PHENIX:2020trf} as well as gluonic exchange currents~\cite{Myhrer:2007cf,Hogaasen:1987nj,Myhrer:2009uq} also contribute to the explanation. 

The final aspect of this problem is that the original naive expectations of the fraction of proton spin carried by quarks ignored the effect of QCD evolution. Even if the gluons carry very little spin at a low scale, relevant to a quark model, this fraction increases logarithmically as the momentum scale rises; albeit with a corresponding increase in gluon orbital angular momentum of opposite sign. Once one combines the effect of the pion field, the gluonic exchange current correction and QCD evolution~\cite{Thomas:2008ga}, the main features of the spin and orbital angular momentum distributions on the quarks in the proton are well reproduced. 

\section{Concluding Remarks}
In this brief review, we have seen that chiral symmetry plays a key role in creating the intricate structure of strongly interacting particles. Future work at facilities such as JLab 12 GeV~\cite{Arrington:2021alx} and the electron-ion collider~\cite{Willeke:2021ymc,AbdulKhalek:2021gbh,Anderle:2021wcy} will probe this structure with ever-increasing precision and in new kinematic conditions. The new information on the origin of the mass and spin of the nucleon~\cite{Ji:1995sv,Balitsky:1997rs,Ji:1996ek}, as well as its three-dimensional structure~\cite{Ji:1998pc,Belitsky:2005qn,Belitsky:2012ch,Accardi:2012qut}, will provide fascinating new insights. However, we can be certain that the interpretation of this new information will continue to depend heavily on chiral symmetry.

\section*{Acknowledgements}
There are many colleagues who have contributed to my understanding of the material described here over many years. To mention a large number would risk insulting others by omission. However, I would especially like to acknowledge the collaborations with Gerry Miller and Serge Th/'eberge in the development of the cloudy bag model and Derek Leinweber in the extraction of new physical results using data from lattice QCD. This work has been supported by the University of Adelaide and by the Australian Research Council through Discovery Project DP150103101, an Australian Laureate Fellowship FL0992247 and through the ARC Centre of Excellence for Dark Matter Particle Physics, CE200100008.

\bibliography{bibliography}

%

\end{document}